\documentclass[aps,pre,showpacs,twocolumn]{revtex4}
\usepackage{graphics}
\usepackage[dvips]{color}
\usepackage[T1]{fontenc}
\usepackage[latin1]{inputenc}
\usepackage{graphicx}
\usepackage{amssymb}
\usepackage{rotating}
\usepackage{epsfig}

\input epsf.sty

\newcommand{\sts}{\scriptsize}

\newcommand{\mb}{\mbox}

\begin{document}

\title{Coarse-grained lattice model for investigating the role of
  cooperativity in molecular recognition}
\author{Hans Behringer, Andreas Degenhard, Friederike Schmid}

\affiliation{Fakult\"at f\"ur Physik, Universit\"at Bielefeld, D -- 33615
Bielefeld, Germany}

\begin{abstract}

\noindent Equilibrium aspects of molecular recognition of rigid
biomolecules are investigated using
coarse-grained lattice models. The analysis is carried out in two stages. First an
ensemble of probe molecules is designed with respect to the target
biomolecule. The recognition ability of the probe ensemble is 
then investigated by calculating the free energy of association. The
influence of cooperative and anti-cooperative effects accompanying the
association of the target and  probe molecules is
studied. Numerical findings are presented and compared to analytical
results which can be obtained in the limit of dominating cooperativity
and in the mean-field formulation of the models.

\end{abstract}


\maketitle

\section{Introduction}

Molecular recognition is the ability of a biomolecule to interact
preferentially with a particular target molecule among a vast variety
of different but almost identically looking rival molecules.  Examples
of specific recognition processes comprise enzyme-substrate binding,
antibody-antigen binding, protein-receptor interactions or
cell-mediated recognition \cite{Alberts_1994,Kleanthous_2000}.
Molecular recognition is essential for biological systems such as the
immune system to work efficiently.  Whereas macromolecules are held
together by covalent bonds the recognition process is governed by
specific non-covalent interactions such as ionic binding, the van der
Waals interaction, the formation of hydrogen bonds and hydrophobicity
\cite{Sneppen_2005}. In an aqueous environment those non-covalent
bonds contribute an energy of the order of 1-2 kcal/mole with the
relatively strong hydrogen bonds sometimes contributing up to 8-10
kcal/mole \cite{Delaage_1991}.  The non-covalent bonds are thus only
slightly stronger than the thermal energy $k_{\mb{\tiny
    B}}T_{\mb{\tiny room}} \simeq 0.62$ kcal/mole at physiological
conditions and therefore the specificity of biomolecule recognition is
only achieved if a large number of functional groups of the two
molecules to recognise each other precisely match and thus a large
number of non-covalent bonds can be formed \cite{Pauling_1940}. The
binding sites of the two molecules are said to be complementary to
each other. This view of molecular recognition for inflexible
macromolecules is sometimes called ``lock-and-key'' mechanism
\cite{Fischer_1894}. However there are notable recognition processes
that involve flexible biomolecules \cite{Peppas_2002}. The matching of
a large number of functional groups is then achieved by a
conformational change giving rise to large entropic contributions
(so-called ``induced fit'' scheme) \cite{Koshland_1958}.  In addition
to short-range interactions ensuring the stability of the complex for
a sufficiently large time long-range electrostatic interactions are
believed to pre-orient the biomolecules so that the probability of the
contact of the complementary patches on the two molecules upon
collision is increased \cite{Janin_2000,Wodak_2003}.

An understanding of the principles of recognition processes between
biomolecules is not only important from a scientific point of view but
also for biotechnological and biomedical applications. The knowledge
of these principles is a necessary input for the design of synthetic
heteropolymers with molecular recognition ability so that they can
interact with a biological environment, i.e. biomolecules, cells and
tissues, in a programmable way (see e.g. the review by
\cite{Peppas_2002}).

In recent years much effort has been spent to investigate the
structural basis for the recognition of two rigid proteins
\cite{Janin_1990,Jones_1996,Janin_2000, Jones_2000,
  Kleanthous_2000,Chakrabarti_2002}. In particular the recognition
sites of the two proteins in contact have been analysed.  The
recognition site on a protein basically consists of the residues,
i.\,e. amino acids, which interact with residues of the other
proteins. It is found to be made up of largely hydrophobic residues so
that its hydrophobicity is comparable with that of the interior of the
protein.  For the development of idealised coarse-grained models it is
therefore assumed that hydrophobicity plays a major role in
recognition processes. Consequently the residue interactions in the
idealised models investigated in this article are assumed to be purely
of hydrophobic nature.

The investigation of the underlying mechanisms of molecular
recognition processes from a physical point of view has recently
gained growing interest. In particular the question of the specificity
of recognition processes has been addressed by methods of statistical
mechanics
\cite{Lancet_1993,Janin_1986,Rosenwald_2002,Wang_2003,Polotsky_2004a,Polotsky_2004b,Bogner_2004,Bernauer_2005,Wang_2006,Behringer_etal_2007,Baake_2007}.
Nevertheless, the view of specificity, which is basically the
occurrence of a preferential binding of the recognition agents in the
presence of a diversity of rival molecules, remains yet incomplete
(the introductory remarks in \cite{Hippel_1986} about the diversity of
definitions of specificity found in the literature still apply).

In this article we develop coarse-grained lattice models for the
investigation of the principles of molecular recognition processes.
Our approach, which is described in section \ref{kap:general},
consists of two stages: In a first step a design of probe molecules is
carried out. This step mimics the design in biotechnological
applications or the evolution in nature. In a second step the
recognition ability is calculated by considering the free energy of
association of the probe molecules with the target and a structurally
different rival molecule. This general approach is illustrated for a
modified hydrophobic-polar (HP) model in section \ref {kap:hp}. In
section \ref{kap:coop} the modified HP model is extended to take
cooperative effects on a residue-specific level into account. The
resulting model is investigated in its mean-field formulation and in
the limiting case of dominant cooperativity which can be tackled
analytically. In addition the model is investigated numerically for
the case where the contributions of the direct residue-residue
interactions and the indirect cooperative interactions are of same
order. The findings are compared to the limiting case of dominating
cooperativity and to the mean field findings.  In section
\ref{kap:alter_coop} another possible way to incorporate cooperative
effects is analysed. The article closes with a conclusion and an
outlook (section \ref{kap:conclusion}). Among other findings some of
the numerical results have been published recently in a separate
Letter \cite{Behringer_etal_2006}.

\section{General approach}

\label{kap:general}

In this article we study coarse-grained models for the recognition of
two rigid proteins. Under physiological conditions the complex of the
proteins is stabilised by non-covalent interactions across its
interface.  The binding of the proteins is accompanied by a decrease
of entropy due to immobilising translational, rotational and
conformational degrees of freedom.  The gain in energy on forming the
complex has thus to be strong enough to overcome these entropic costs.
In our model the proteins are considered to be rigid so that
conformational changes of the backbone of the proteins need not be
taken into account. This assumption is fulfilled for a large variety
of real protein-protein associations, however, even for the
association of two rigid proteins minor rearrangements of the side
chains of the amino acids will occur (e.\,g. \cite{Janin_1990}).

The energetics at the contact interface of the complex can be
formulated in a coarse-grained way where coarse-graining is adopted
for both the structural properties of the recognition sites of the
involved biomolecules and for the interaction between two residues
\cite{Behringer_etal_2006}.  Consider a recognition site of $N$
residues on both proteins. For simplicity it is assumed that the two
recognition sites contain the same number of residues and that
precisely two residues match respectively in the interface. Notice
that a recognition site found in natural protein-protein complexes
contains typically of the order of $30$ residues
\cite{Janin_2000,Jones_2000}.  The chemical structure of the
recognition site of the protein to be recognised, which is called {\it
  target} molecule in the following, is characterised in a
coarse-grained approach by a discrete variable $\sigma = (\sigma_1,
\ldots, \sigma_N)$ where the value of $\sigma_i$ specifies the type of
the residue at positions $i$, $i=1, \ldots, N$, on the recognition
site. Similarly, the types of the residues of the recognition site of
the other protein which recognises the target are specified by a
second variable $\theta = (\theta_i, \ldots, \theta_N)$. In the
following this second biomolecule is called {\it probe} molecule.  On
a coarse-grained level, the interaction of the functional groups
across the interface is described by a Hamiltonian
$\mathcal{H}(\sigma, \theta; S)$ where we incorporate an additional
interaction variable $S = (S_1, \ldots, S_N)$. The variable $S_i$
takes the quality of the contact of the residues of the two proteins
at position $i$ into account, where a good contact leads to a
favourable contribution to binding and a bad one only to a small
contribution. A good contact may imply, for example, that the distance
between the two residues is small or the polar moments of residues are
appropriately aligned to each other.  A steric hindrance on the other
hand may result in a large distance between the residues and
consequently one has a bad contact. The variable $S$ therefore models
effects that are related to minor rearrangements of the side-chains of
the amino acids when the two proteins form a complex.

Along these general lines a first model, namely a modification of the
hydrophobic-polar (HP) model, can be formulated. In the HP-model only
two different types of residues are distinguished, namely hydrophobic
(H) and polar (P), i.\,e. hydrophilic, ones, so that the variables
$\sigma$ and $\theta$ specify the degree of hydrophobicity of the
residues.  This restriction to the hydrophobic interaction is
motivated by the observation that the hydrophobicity is a major
property that discriminates the recognition site from other patches on
the surface of a protein. Hydrophobic residues are described by the
variable $\sigma_i = +1$ and polar residues by $\sigma_i = -1$.  The
Hamiltonian is then given by
\begin{equation}
\label{eq:HP-hamiltonian}
  \mathcal{H}_{\mb{\sts HP}}(\sigma,\theta;S) =- \varepsilon \sum_{i=1}^N\frac{1+S_i}{2} \sigma_i\theta_i
\end{equation}  
where the sum extents over the $N$ positions of the residues of the
recognition site and the interaction constant $\varepsilon$ is
positive. It is typically of the order of $\varepsilon \simeq 1$
kcal/mole for hydrophobic interactions \cite{Sneppen_2005}. The
product $-\varepsilon\sigma_i\theta_i$ describes the mutual
interaction of the residues in contact across the interface. The
additional variable $S_i$ can take on the values $\pm 1$. Thus for
$S_i = +1$ one has a good contact leading to a non-zero contribution
to the total interaction energy, for $S_i = -1$, on the other hand,
one has a bad contact and no energy contribution. Notice that a good
contact not necessarily leads to a favourable energy contribution.
Note also that the original HP-model, which has been introduced to
study the protein folding problem \cite{Dill_1985}, does not contain
an additional variable $S$ to model the quality of the contact.

The grouping of the 20 natural amino acids into classes of
characteristic types is very important for the development of minimal
models for the study of protein interactions. The reduction to a
hydrophobic and a polar type and thus the use of an Ising-like model
Hamiltonian such as (\ref{eq:HP-hamiltonian}) on a coarse-grained
level is also justified by the findings in \cite{Li1997}. In this work
Li et al. applied an eigenvalue decomposition to the Miyazawa-Jernigan
matrix of inter-residue contact energies of amino acids. They found
that the interaction matrix can be parameterised by an Ising-like
model where the ``spin variable'' can take on different discrete
values. As these values show a bimodal distribution the
reparameterisation basically reduces to the Ising model where the two
possible values of the ``spins'' describe hydrophobic and polar
residues. Introducing the additional variable $S$ for the
rearrangement of the amino acid side chains we end up with Hamiltonian
(\ref{eq:HP-hamiltonian}). Suggested by experimental observations the
grouping of the amino acids into five characteristic groups is also
widely discussed \cite{WangWang_1999,Cieplak_2001}. The reduction in
\cite{Cieplak_2001}, for example, uses a distance-based clustering
applied to the Miyazawa-Jernigan matrix.  The resulting grouping
reproduces the statistical and kinetic features of well-designed
sequences in the protein-folding problem. The grouping into five
different characteristic types in these approaches points at possible
extensions of our model for the contact interaction. In this work,
however, we restrict ourselves only to hydrophic and polar residue
types.

To study the recognition process between the two biomolecules we adopt
a two-stage approach. First the structure of the recognition site of
the target molecule is fixed to a certain sequence $\sigma^{(0)} =
(\sigma_1^{(0)}, \ldots, \sigma_N^{(0)})$ of residues. Then this
structure is learned by the probe with respect to some learning rules
under conditions that are specified by a parameter $\beta_{\mb{\tiny
    D}}$. This leads to an ensemble of probe molecules of sequences
$\theta$ at their recognition sites with a probability
$P(\theta|\sigma^{(0)})$ depending on the initially fixed target
structure. To illustrate this a bit further consider a design step
where learning is done just by thermal equilibration. The probability
distribution is then technically given by the canonical Boltzmann
distribution
\begin{equation}
P(\theta | \sigma^{(0)}) = \frac{1}{Z_{\sts \mb{\tiny D}}} \sum_{S}
    \exp\left(-\beta_{\sts\mb{\tiny D}} \mathcal{H}(\sigma^{(0)}, \theta;S)
    \right)
\end{equation}
where the normalisation $Z_{\sts \mb{\tiny D}}$ is the usual canonical
partition function. The design temperature $\beta_{\sts\mb{\tiny D}}$
acts as a Lagrange multiplier that fixes the average energy, however,
the parameter $\beta_{\sts\mb{\tiny D}}$ may also be interpreted to
describe more generally the conditions under which the design has been
carried out. This first design step is introduced to mimic the design
in biotechnological applications or the process of evolution in nature
\cite{anmerk}. Note that in some studies of the protein folding
problem \cite{Pande_1997, Pande_2000} and the adsorption of polymers
on structured surfaces \cite{Jayaraman_2005} a similar design step has
been incorporated.

In the second step the recognition ability of the designed probe
ensemble of structures $\theta$ is tested. To this end the ensemble is
brought into interaction with both the picked target structure
$\sigma^{(0)}$ and a competing (different) structure $\sigma^{(1)}$ at
some inverse temperature ${\beta}$ which in general is different from
the design temperature $\beta_{\mb{\tiny D}}$. The free energy of the
probe system interacting with the structure $\sigma^{(\alpha)}$,
$\alpha = 0, 1$, is then
\begin{equation}
F^{(\alpha)} =
\sum_\theta F(\theta|\sigma^{(\alpha)})P(\theta|\sigma^{(0)})
\end{equation}
where $F(\theta|\sigma^{(\alpha)})$ is the thermal free energy for the interaction
between $\sigma^{(\alpha)}$ and a fixed probe sequence $\theta$ and an
average over the structures in the probe ensemble is carried out. The
free energy $F(\theta|\sigma^{(\alpha)})$ is given by
\begin{equation}
F(\theta|\sigma^{(\alpha)}) = -\frac{1}{\beta} \ln \sum_{S} \exp\left(- \beta \mathcal{H}(\sigma^{(\alpha)},
  \theta;S) \right).
\end{equation}
The target with the structure $\sigma^{(0)}$ at its recognition site
is recognised by the probe if the associated free energy $F^{(0)}$ is
lower than the free energy $F^{(1)}$ for the interaction with the
competing structure $\sigma^{(1)}$, i.\,e. in a mixture of equally
many $\sigma^{(0)}$ and $\sigma^{(1)}$ molecules the probe molecules
preferentially bind to the original target.  This is signalled by a
negative free energy difference $\Delta F(\sigma^{(0)},\sigma^{(1)}) =
F^{(0)} - F^{(1)}$.  Thus the specificity of the recognition process
is related to the difference between the free energy of association
for the competing molecules. For given structures $\sigma^{(0)}$ and
$\sigma^{(1)}$ one can introduce a suitable measure $Q$ for the
structural similarity of the target and the rival biomolecule.
Carrying out an average over all target and rival structures that are
compatible with the specified similarity $Q$ one can compute the
averaged free energy difference of association $\Delta F(Q)$ as a
function of the similarity between the target and the rival and
therefore investigate the overall recognition ability of the model
(see section \ref{kap:hp} below for the HP-model). Note that in our
approach the mechanism which brings the two interacting molecules, in
particular the two recognition sites, into contact is not taken into
account, i.\,e., only equilibrium aspects are considered.

In principle interactions of the residues which do not belong to the
recognition sites with solvent molecules have to be considered as
well. Solvation effects at the recognition sites and the associated
entropy changes are also important for the association process of
biomolecules \cite{Gilson_1997,Jackson_2006}.  In the coarse-grained
model, however, it is assumed that all these contributions are of the same
size for all proteins under consideration.  Note also that solvation
effects are already partially contained in HP-models.  In addition the
entropic contributions due to a reduction of the translational and
rotational degrees of freedom upon forming a complex can be assumed to
cancel out in the free energy difference $\Delta F$ in a first
approximation.  This requires at least that the two competing proteins
are of comparable shape and size.

In this work we assume that the proteins have the same number of
residues at the interface. However, many protein-protein interfaces
are curved with different numbers of residues on the two proteins
\cite{Wodak_2003}.  Nevertheless, we expect our
assumption not to be crucial within the above simplified
coarse-grained view, at least in a first approximation. As our model
characterises the residues only with respect to their hydrophobicity
one can partition the interface into $N$ contacts and attribute
hydrophobicities to the patches on the two proteins that contribute to
a particular contact. Then one ends up again with our Hamiltonian
(\ref{eq:HP-hamiltonian}). For approaches where the residue type is
determined by additional features apart form hydrophobicity
correlations between neighbouring patches might occur so that our
assumption may become questionable.

\section{Application to a modified hydrophobic-polar model}
\label{kap:hp}

The modified HP-model of the previous section, can again serve as an
illustration of the two-state approach for investigating molecular
recognition processes. As (\ref{eq:HP-hamiltonian}) does not involve
any interaction between neighbouring residues of the recognition site
of a protein, the two-stage approach can be worked out exactly.

\paragraph{Design by equilibration}

For the HP-model
the design governed by thermal equilibration leads to the conditional probability
\begin{equation}
\label{eq:hp_model_design}
P(\theta|\sigma^{(0)}) = \frac{\exp\left( \frac{\varepsilon\beta_{\mb{\tiny
          D}}}{2} \sum\limits_i \theta_i \sigma^{(0)}_i\right)}{\left (
          2\cosh\left(\frac{\varepsilon\beta_{\mb{\tiny
          D}}}{2}\right)\right)^N}
\end{equation}
of the structure $\theta$ at the recognition site of the probe
molecule. As mentioned in the previous section the design temperature
$\beta_{\mb{\tiny D}}$ may be interpreted to characterise the
conditions under which the design has been carried out. This can be
illustrated using the present example of the HP-model. In the HP-model the
value of $\sigma_i$ or $\theta_i$, respectively, basically
specifies the hydrophobicity of the residue at position $i$. The
total hydrophobicity of the recognition site of the target molecule is then
$H^{(0)} = \sum_i \sigma^{(0)}_i$. From relation
(\ref{eq:hp_model_design}) one can calculate the average
hydrophobicity $\left<H_{\mb{\tiny D}}\right>$ of probe structures:
\begin{equation}
  \left<H_{\mb{\tiny D}}\right> = \sum_k \sum_{\theta}\theta_k
  P(\theta|\sigma^{(0)}) = H^{(0)} \tanh\left(\frac{\varepsilon\beta_{\mb{\tiny D}}}{2}\right).
\end{equation} 
Thus the Lagrange parameter $\beta_{\mb{\tiny D}}$ can be used
to fix the average hydrophobicity of the designed probe ensemble which
is achieved by controlling the supply of hydrophobic residues during the
design procedure.

The probability distribution (\ref{eq:hp_model_design}) for the
designed structures $\theta$ explicitly depends on the structure
$\sigma^{(0)}$ of the recognition site of the fixed target molecule.
For the HP-model a design under ideal conditions, i.\,e. $1/\beta_{\mb{\tiny D}} = 0$, the structure $\theta$ would
simply be a copy of $\sigma^{(0)}$. However, for
non-ideal conditions with $\beta_{\mb{\tiny D}} < \infty$ ``defects''
appear in the design procedure and the obtained structure $\theta$
deviates on average from $\sigma^{(0)}$. This deviation can be
quantified by the complementarity parameter
\begin{equation}
\mathcal{K}(\theta, \sigma^{(0)}) = \sum\limits_i \theta_i \sigma^{(0)}_i.
\end{equation}
The possible values of $\mathcal{K}$ range from $-N$ to $N$ in even
steps. A value $\mathcal{K}(\theta, \sigma^{(0)})$ close to $N$ means
a high complementarity and the interaction of the probe structure
$\theta$ with $ \sigma^{(0)}$ can lead to a large enough energy
decrease so that a complex can be formed. On the other hand a value of
$\mathcal{K}(\theta, \sigma^{(0)})$ much less than $N$ signals a poor
match between the two recognition sites and therefore it is unlikely
that a complex is stabilised.

The probability distribution $P(\theta|\sigma^{(0)})$ can be converted
to a distribution function for the complementarity leading to the
probability
\begin{eqnarray}
  P(K) & =& \sum_\theta P(\theta|\sigma^{(0)})
  \delta_{\mathcal{K}(\theta, \sigma^{(0)}),K}
  \\
 &=& {N \choose
  \frac{1}{2}(N+K)}
\frac{ \exp \left( 
  \frac{\varepsilon\beta_{\mb{\tiny
          D}}}{2} K\right)}{\left (
          2\cosh\left(\frac{\varepsilon\beta_{\mb{\tiny
          D}}}{2}\right)\right)^N}
\end{eqnarray} 
to have a complementarity parameter $K$ in the designed ensemble. The
quality of the design can now be measured by the average
complementarity of the designed structures $\theta$ which is given by
\begin{equation}
  \left< K \right> = \sum_{K} K P(K) = N \tanh\left(
  \frac{\varepsilon\beta_{\mb{\tiny
          D}}}{2} \right)
\end{equation}
for the modified HP-model. For large $\beta_{\mb{\tiny D}}$ one gets a
probe ensemble which is fairly complementary to the fixed target
structure.  Thus large values of $\beta_{\mb{\tiny D}}$ correspond to
good design conditions, an observation which can already be deduced
from the interpretation of $\beta_{\mb{\tiny D}}$ as an inverse
temperature. In the hydrophobicity interpretation discussed above
large values of $\beta_{\mb{\tiny D}}$ signify comparable
hydrophobicities of the target and the probe molecule.

\paragraph{Recognition ability}

The recognition ability of the probe molecules is tested by comparing
the free energy of association with the target structure
$\sigma^{(0)}$ and a competing molecule with the structure
$\sigma^{(1)}$ at its recognition site. For the HP-model
(\ref{eq:HP-hamiltonian}) with its two different types of residues,
one can introduce the similarity parameter
\begin{equation}
  Q = \sum_i \sigma^{(0)}_i\sigma^{(1)}_i.
\end{equation} 
For $Q$ close to its maximum value $N$ the competing molecule has a
recognition site that is almost identical to the one of the target
molecule. In terms of the similarity parameter $Q$ the free energy
difference is given by 
\begin{equation}
\label{eq:deltafreinhp}
  \Delta F(Q) = -\frac{1}{2}\varepsilon N
  \tanh\left(\frac{\varepsilon \beta_{\mb{\tiny D}}}{2}\right) (N-Q).
\end{equation} 
The free energy difference is always negative as soon as the
recognition site of the competing molecule is not identical to the one
on the target molecule.  In equilibrium the probe molecule therefore
binds preferentially to the target molecule and thus the target
molecule is specifically recognised. The difference in free energy
increases for an decreasing similarity parameter $Q$. Note also that
the slope of the free energy difference depends only on the conditions
under which the design of the probe molecules has been carried out.

\section{Role of cooperativity in molecular recognition}
\label{kap:coop}

Cooperative effects play an essential role in many biological
processes such as the catalysis of biochemical reactions by enzymes.
Cooperativity is presumably also very important for molecular
recognition processes \cite{Cera_1998}. In general cooperativity means
that the binding strength of two residues depends on the binding
interactions in the neighbourhood of the two residues in contact. Thus
the energetic properties of residues when interacting with other
residues cannot be inferred by considering them isolated from the
local environment. This has implicitly been done, however, in the
modified HP model (\ref{eq:HP-hamiltonian}) where the interaction
constant $\varepsilon$ has been attributed to the residue-residue
interaction independently of the corresponding local environment.

In this section the modified HP-model of the preceeding section is
extended to incorporate the effect of cooperative interactions on
molecular recognition. Note that in reference \cite{Cera_1998} it has
been argued that cooperativity should be incorporated on a
residue-specific level.

During the association process rearrangements of the amino acid side
chains are observed. Thus in the idealised model applied in this work
cooperative effects stem from the behaviour of the variables $S_i$.  A
possible extension of the modified HP-model which takes cooperative
interactions into account is given by
\begin{equation}
\label{eq:HP-coop}
\mathcal{H}(\sigma, \theta;S) = - \varepsilon\sum_{i=1}^{N} \frac{1+S_i}{2} \sigma_i \theta_i - J
\sum_{\left<ij\right>} S_i S_j.
\end{equation}
The first sum describes again the hydrophobic interaction whereas the
second sum represents the additional cooperative interaction. It
extends over the neighbour positions of the residue at position $i$.
For a fixed $i$ on a square-lattice the sum includes therefore four
terms.  The interaction coefficient $J$ is positive for cooperative
interactions and negative for anti-cooperative interactions. To get an
impression of its effect consider the design step. Suppose that at
position $i$ one has a hydrophobic residue on the target molecule.
Then the first term in the HP-Hamiltonian (\ref{eq:HP-coop}) favours
that a hydrophobic residue adsorbs there with a good contact $S_i =
+1$ on average.  Suppose now that on one of the neighbouring positions
$j$ of $i$ on the target one has again a hydrophobic residue. If a
hydrophobic residue gets adsorbed at the corresponding position of the
probe structure a good contact with $S_j = +1$ is favoured by the
hydrophobic interaction term in (\ref{eq:HP-coop}). But then the
second cooperative term leads to an additional energy decrease for
$J>0$. If on the other hand a polar residue shows up at the position
$j$ on the probe molecule the hydrophobic contribution in
(\ref{eq:HP-coop}) tries to avoid a contact, i.\,e.  $S_j = -1$, on
average, which then leads to an unfavourable energy increase due to
the cooperative term. The quality of a contact thus couples to the
quality of the contacts in the neighbourhood of a residue. For a
positive constant $J$ the cooperativity is therefore expected to
enhance the fit of the molecules at the interface resulting in an
increased average complementarity compared to an interaction without
cooperativity.  Similarly, one expects an increase in the recognition
specificity. In the subsequent paragraphs these suggestions are
investigated for cooperative interactions.

Note that $\theta_i$ (and thus the product $\sigma_i\theta_i$) in
Hamiltonian (\ref{eq:HP-coop}) is a random variable whose distribution
is obtained by the design step. The energy function (\ref{eq:HP-coop})
describes therefore a random field Ising model. Contrary to the models
mostly investigated in the literature (e.\,g.
\cite{Dotsenko_2001,Schneider_1977}) the distribution function of the
random variable $\sigma_i\theta_i$ is not symmetric with respect to a
sign-reflection.

\subsection{Limiting case of dominant cooperativity}
\label{kap:Jgross}

The case where the cooperative contribution to the total energy
dominates can be investigated analytically. Consider the situation
where $J \gg N\varepsilon$. The cooperative term $- J
\sum_{\left<ij\right>} S_i S_j$ in the Hamiltonian (\ref{eq:HP-coop})
has discrete energy levels $-4NJ, -4(N-1)J, \ldots, +4NJ$ for a
recognition site with a rectangular geometry where each residue has
four neighbours. The hydrophobic interaction term $-\varepsilon \sum_i
\frac{1+S_i}{2} \sigma_i \theta_i$ has also discrete levels ranging
from $-N\varepsilon$ to $+N\varepsilon$. For the above assumption $J
\gg N\varepsilon$ the global rough structure of the spectrum of the
Hamiltonian ({\ref{eq:HP-coop}}) is basically determined by the
cooperative contributions.  The hydrophobic interaction of the
residues in contact introduces only small variations about the main
energy levels with two adjacent ones being separated by an amount of
$4J$. For a small temperature, i.\,e.  a large $\beta$, the
statistical behaviour is dominated by the twofold degenerate lowest
energy state of the cooperative interaction term with all $S_i$ being
either in the state $+1$ or in the state $-1$. Due to this reduction
of the phase space of possible $S$ configurations the two-stage
approach can be worked out analytically.  The dominance of the
cooperative term leads to the new effective Hamiltonian
\begin{equation}
  \label{efffuerJgross}
  \mathcal{H} \stackrel{J\gg \varepsilon N}{\sim} - \frac{1+s}{2} \varepsilon\sum_{i=1}^{N} \sigma_i \theta_i - 4NJ
\end{equation}
where the scalar variable $s$ can have the values $\pm 1$. The design step
now yields the probability distribution
\begin{equation}
  P(\theta|\sigma^{(0)}) = \frac{ 1 + \exp\left( {\varepsilon\beta_{\mb{\tiny
            D}}} \sum\limits_i \theta_i \sigma^{(0)}_i\right)}{ 2^N + \left (
      2\cosh\left({\varepsilon \beta_{\mb{\tiny
              D}}}\right)\right)^N}
\end{equation}
for the structure of the recognition site of the probe molecules. The
corresponding distribution of 
the complementarity between the structures $\sigma^{(0)}$ and $\theta$ is 
\begin{equation}
  \label{eq:pvonkmitj}
P(K) = {N \choose \frac{1}{2}(N+K)}\frac{ 1 + \exp\left( {\varepsilon\beta_{\mb{\tiny
            D}}} K\right)}{ 2^N + \left (
      2\cosh\left({\varepsilon \beta_{\mb{\tiny
              D}}}\right)\right)^N}.
\end{equation}
The distribution function for the complementarity parameter $K$ can
again be used to calculate the average complementarity of the designed
molecules. For large $N$ (for which the term $2^N$ in the denominator
of (\ref{eq:pvonkmitj}) can be neglected as long as $\beta_{\mb{\tiny
          D}} \neq 0$) one obtains 
\begin{equation}
\label{eq:kompmitj}
  \left< K \right> \stackrel{J\gg\varepsilon N}{\sim}  N \tanh\left(
  {\varepsilon \beta_{\mb{\tiny
          D}}} \right).  
\end{equation}
In the situation of a dominating cooperative interaction the average
complementarity is increased compared to the case where cooperativity
is absent. This suggests that values of the
cooperativity interaction constant $J$  comparable to the size of the
hydrophobic interaction constant $\varepsilon$ might also
enhance the quality of the design step. This question is investigated
numerically in the subsequent paragraph \ref{kap:numJ}.

In the second step the designed probe ensemble interacts with the chosen
target structure $\sigma^{(0)}$ and a competitive one
$\sigma^{(1)}$. The associated free energy averaged with respect to
the distribution of the structures $\theta$ of the probe molecules is in
general given by
\begin{eqnarray}
  F^{(\alpha)} &=& -\frac{1}{\beta} \sum_{\theta} \ln\left(  1 + 
\exp\left( \varepsilon\beta \sum\limits_i \theta_i
  \sigma^{(\alpha)}_i\right)\right)
\\ && \times
\frac{ 1 + \exp\left( {\varepsilon\beta_{\mb{\tiny
            D}}} \sum\limits_i \theta_i \sigma^{(0)}_i\right)}{ 2^N + \left (
      2\cosh\left({\varepsilon \beta_{\mb{\tiny
              D}}}\right)\right)^N}.
\end{eqnarray} 
In case of a large number of residues $N\gg 1$ again further progress can be
made analytically. Consider first the free energy of association of the system with
the fixed target structure. In this case the sum over the possible
structures of the designed probe molecules can be converted into a sum
over the complementarity parameter $K$:
\begin{equation}
\label{eq:f0mitk}
  F^{(0)} = -\frac{1}{\beta} \sum_K \ln\left( 1 + \exp(\varepsilon\beta K)\right) P(K).
\end{equation} 
The dominant contributions to this sum arise from the values of $K$
close to the maximum of the distribution $P(K)$. For suitably large
$\beta_{\mb{\tiny D}}$ this maximum, however, occurs for $K \sim
\mathcal{O}(N)$ and thus it is large as well (compare relation
(\ref{eq:kompmitj})).  Therefore, in the limit $N \gg 1$ one can use
the replacements $1 + \exp(\beta_{\mb{\tiny D}}\varepsilon K) \approx
\exp(\beta_{\mb{\tiny D}}\varepsilon K)$ and $\ln(1 +
\exp(\beta\varepsilon K)) \approx \beta\varepsilon K$. Using these
approximations the summation in
(\ref{eq:f0mitk}) leads to 
\begin{equation}
  F^{(0)} \stackrel{N \gg 1}{\sim} -\varepsilon \sum_K KP(K) =
  -\varepsilon \left<K\right> \stackrel{(\ref{eq:kompmitj})}{=} -\varepsilon N \tanh(\varepsilon \beta_{\mb{\tiny D}}).
\end{equation}

A similar conversion cannot be applied to the summation over the
designed molecules in the calculation of $F^{(1)}$ as both
$\theta_i\sigma^{(0)}_i$ and $\theta_i\sigma^{(1)}_i$ terms
appear. Defining the auxiliary variables $k_i :=\theta_i\sigma^{(0)}_i$ and $q_i :=
\sigma^{(0)}_i\sigma_i^{(1)}$ and noting that $(\sigma^{(\alpha)}_i)^2 =
1$ the free energy $F^{(1)}$ is explicitly
given by 
\begin{eqnarray}
\label{eq:f1mitk}
F^{(1)} &=& -\frac{1}{\beta} \sum_{k} \ln\left(  1 + 
\exp\left( \varepsilon\beta \sum\limits_i k_iq_i\right)\right)
\\      &&\times     \frac{ 1 + \exp\left( {\varepsilon\beta_{\mb{\tiny
            D}}} \sum\limits_i k_i\right)}{ 2^N + \left (
      2\cosh\left({\varepsilon \beta_{\mb{\tiny
              D}}}\right)\right)^N}.
\end{eqnarray}
The variable $k_i$ specifies the local complementarity between the
target $\sigma^{(0)}$ and a particular probe structure $\theta$.
Using again the observation that the dominant contributions originate
from values of large $K = \sum_i k_i$ one can use again the replacement $1 +
\exp(\beta_{\mb{\tiny D}}\varepsilon K) \approx \exp(\beta_{\mb{\tiny
    D}}\varepsilon K)$. The logarithmic factor in (\ref{eq:f1mitk})
gives large contributions if the majority of the $q_i$ variables is in state
$+1$. Thus, in the limit of $Q = \sum_i q_i \gg 1$ the sum in
(\ref{eq:f1mitk}) can be worked out and one obtains
\begin{equation}
  F^{(1)} \stackrel{N,Q \gg 1}{\sim} -\varepsilon Q
  \tanh(\varepsilon\beta_{\mb{\tiny D}}).
\end{equation} 
The free energy difference in terms of the similarity $Q$ of the 
competing molecules $\sigma^{(0)}$ and $\sigma^{(1)}$ is now given by 
\begin{equation}
\label{eq:ffuerqpos}
  \Delta F \stackrel{N,Q \gg 1}{\sim} -\varepsilon  \tanh(\varepsilon
  \beta_{\mb{\tiny D}}) (N - Q)
\end{equation}
for positive and large $Q$. Again one has a linear dependence in the
vicinity of $Q=N$. This can be compared to the corresponding result
(\ref{eq:deltafreinhp}) for the situation with $J = 0$. The
cooperativity increases the slope of the free energy difference and
thus the recognition ability of the designed probe ensemble is
increased by cooperativity.

In the limit $Q = \sum_i q_i \ll -1$, on the other hand, almost all
$q_i$ take on the value $-1$ and thus $\sum_i k_iq_i$ is close to $-N$
for those $k_i$ leading to the dominant contributions in
(\ref{eq:f1mitk}). One therefore has 
\begin{equation}
  \ln \left( 1 + 
\exp\left( \varepsilon\beta \sum\limits_i k_iq_i\right) \right)
\stackrel{N \gg 1,Q \ll -1}{\sim} \exp\left( -\varepsilon \beta N\right)
\end{equation}
for the logarithmic factor of the dominant terms in (\ref{eq:f1mitk}).
The free energy of association of the probe molecules with the rival
molecule is thus $F^{(1)} \sim \mathcal{O}(\mb{e}^{-N})$
so that
\begin{equation}
\label{eq:ffuerqneg}
  \Delta F \stackrel{N \gg 1,Q \ll -1}{\sim} F^{(0)}=  -\varepsilon N \tanh(\varepsilon
  \beta_{\mb{\tiny D}}).  
\end{equation}
In the limit $Q\ll -1$ the free energy difference is thus independent
of the similarity parameter $Q$ between the target structure and the rival
structure.

For a similarity parameter $|Q|\sim \mathcal{O}(1)$ one expects
deviations form the behaviour for large $|Q|$. For
the free energy difference per residue $\Delta F /N$ as a function of
the similarity per residue $Q/N$, however, the deviations show up for
similarities $Q/N$ of the order of $1/N$. The free energy difference
per residue will thus develop a kink at $Q/N = 0$ in the asymptotic
limit of $N\to \infty$ so that it is given by expression
(\ref{eq:ffuerqpos}) for positive $Q/N$ and by relation
(\ref{eq:ffuerqneg}) for negative $Q/N$. The range of values of the
similarity per residue where deviations between the free energy for a
system with finite $N$ and the asymptotic result show up is shrinking
for increasing $N$.

\subsection{Numerical results for arbitrary cooperativity}
\label{kap:numJ}

The above analysis of the limiting case $J \gg N\varepsilon$ with a dominant cooperative
interaction suggests that cooperativity
enhances the quality of the design step and eventually increases the
recognition ability. In
this section this suggestion is investigated more closely for
cooperativity constants $J$ which are of the order of the interaction
constant $\varepsilon$ of the hydrophobic interaction term in (\ref{eq:HP-coop}).

\paragraph{Design.}

For non-zero, but finite values of $J$ it is not possible any more to solve the
model analytically. Therefore, the two-stage approach has to be carried
out numerically. To this end the density of states for the design step
is calculated as a function of the energy and the complementarity
parameter for a fixed cooperativity $J$. The density of states is 
generally given by
\begin{equation}
  \Omega_J(K;E) = \sum_{\theta,s} \delta_{K, \mathcal{K}(\theta,
  \sigma^{(0)})}\delta_{E, \mathcal{H}(\theta, \sigma^{(0)};S)}
\end{equation}   
for a fixed target structure $\sigma^{(0)}$. The density of states
$\Omega_J(K;E)$ is thus the number of $(\theta,S)$ configurations that
have energy $E$ when interacting with the target and a complementarity
$K$ of the probe molecule $\theta$ to the target recognition site. In
general the density of states depends additionally on the
configuration $\sigma^{(0)}$ of the recognition site of the target
molecule. However, for the HP-model (\ref{eq:HP-coop}) the density of
states has no explicit dependence on $\sigma^{(0)}$ as the variables
$\theta_i$ can be transformed to the auxiliary variables $k_i :=
\sigma^{(0)}_i\theta_i$, which have the same phase space as
$\theta_i$, so that $\sigma^{(0)}$ does not appear any more.

\begin{figure}[h!]
\begin{center}
\includegraphics[scale=0.285,angle=0]{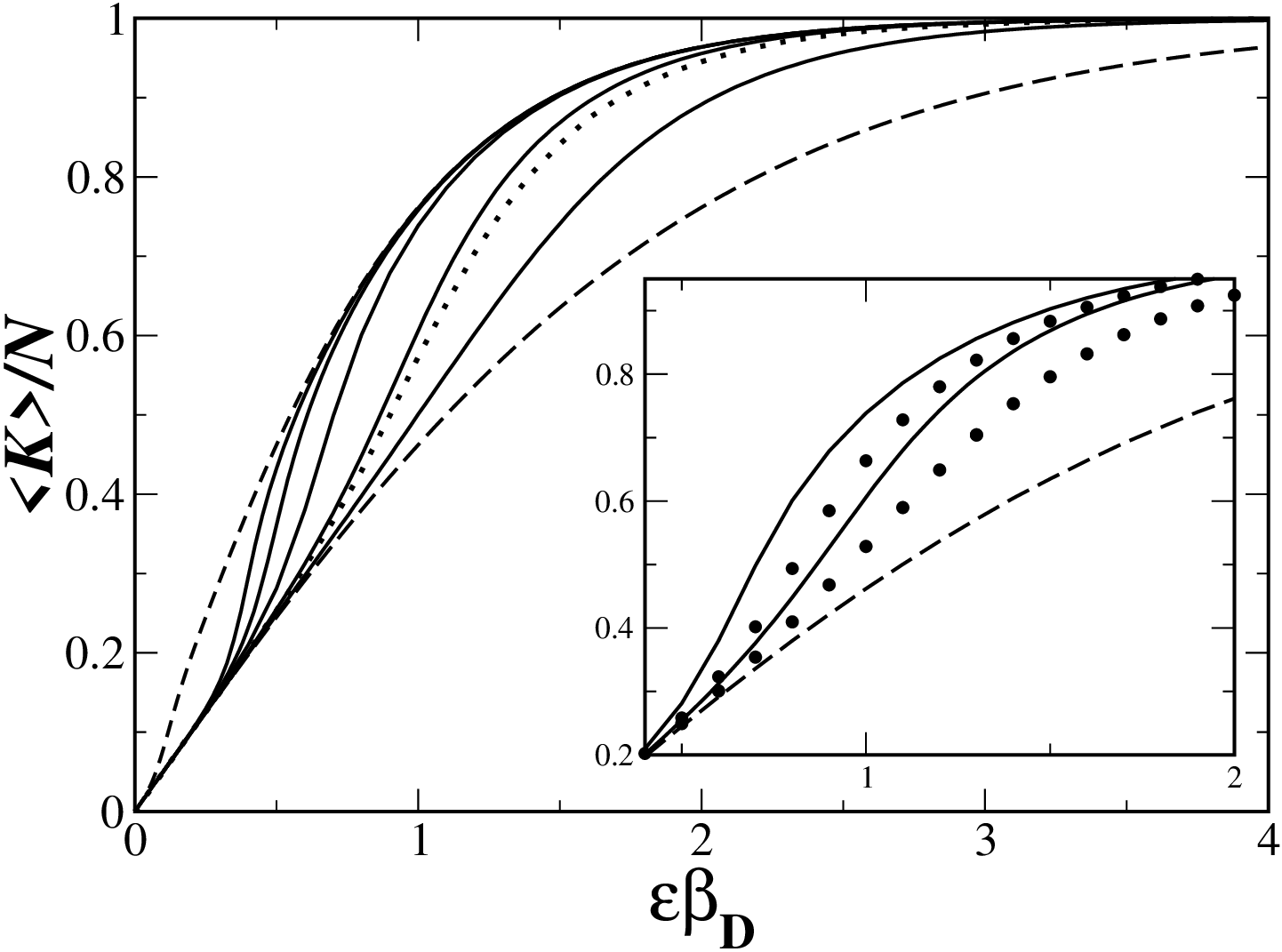}
\caption{\label{bild:coop_comp} Average complementarity per site of the designed ensemble for the
  HP-model (\ref{eq:HP-coop}) for different values of $J$. The lower
  dashed curve corresponds to $J=0$, the upper dashed line represents
  the limiting case $J \to \infty$ (for large $N$). The solid curves
  in between from the bottom up correspond to the values 0.1, 0.25,
  0.5, 0.75 and 1.0 of $J$ in units of $\varepsilon$.  The dotted
  curve shows the result for a system with additional next nearest
  neighbour cooperativity with $J_{\mb{\tiny nnn}}/\varepsilon =
  J_{\mb{\tiny nn}}/\varepsilon = 0.1$.  The inset compares the
  numerical results (solid lines) with the mean field findings of
  section \ref{kap:meanfieldJ} (circles) for cooperativities 0.25 and
  0.5. The dashed curve corresponds again to $J=0$.  }
\end{center}
\end{figure}

The density of states can be calculated directly using efficient Monte
Carlo algorithms \cite{Hueller_2002, Wang_2001,Landau_2004}. In this work the
Wang-Landau algorithm has been applied.  Once the density of states is
known the probability distribution  of the complementarity is basically
obtained by calculating the Laplace transform of $\Omega_J$ giving up
to a normalisation
\begin{equation}
  P_{J}(K; \beta_{\mb{\tiny D}}) \sim \sum_E \Omega_J(K, E)
  \exp(-\beta_{\mb{\tiny D}}E).
\end{equation}
From this distribution function one can calculate the average
complementarity $\left< K \right> (J) =\sum_K P_{J}(K;
\beta_{\mb{\tiny D}})K$ which is shown in figure \ref{bild:coop_comp}.
The calculations have been carried out for a square-lattice geometry
with $N=256$ residues.  We have checked that the curves show only
minor finite-size effects for recognition sites of realistic sizes
with $N\sim \mathcal{O}(30)$ (see \cite{Behringer_etal_2006}). The
qualitative findings discussed in the following are independent of the
number $N$ of residues involved in the interface.
It is visible that cooperativity increases the average complementarity
of the probe molecules for large enough values of $\beta_{\mb{\tiny
    D}}$.  For a parameter value of the order of $\varepsilon
\beta_{\mb{\tiny D}} \simeq 1$ a small change in the cooperativity $J$
leads to a large difference in the average complementarity. Therefore,
small changes in $J$ can have a large impact on the quality of the
design step. As the typical energy $\varepsilon$ of a non-covalent
bond is of the order of 1 kcal/mole this regime corresponds indeed to
physiological conditions.

The Hamiltonian (\ref{eq:HP-coop}) contains a cooperative term where
the quality of the contact couples to the contact variable at the
neighbouring sites. This limitation to nearest neighbour interactions
can be relaxed by allowing additional couplings to sites that are
further away. As long as the range of the cooperative coupling is
finite, however, we expect, that the average complementarity
$\left<K\right>$ is qualitatively similar as for the model
(\ref{eq:HP-coop}). For the system with nearest and next nearest
neighbour interactions (with the same constant $J$) the case of
dominant cooperativity can be treated as above (section
\ref{kap:Jgross}) leading to the same effective Hamiltonian
(\ref{efffuerJgross}) with the irrelevant constant replaced by $-8NJ$.
So the same limiting curves for $\left<K\right>$ as well as $\Delta F$
result. However, the additional interactions have the consequence that
the maximum effect of cooperativity will already show up for smaller
values of $J$. This is shown in figure \ref{bild:coop_comp} for the
model with additional next nearest neighbour cooperativity.

\begin{figure}[h!]
\begin{center}
\includegraphics[scale=0.285,angle=0]{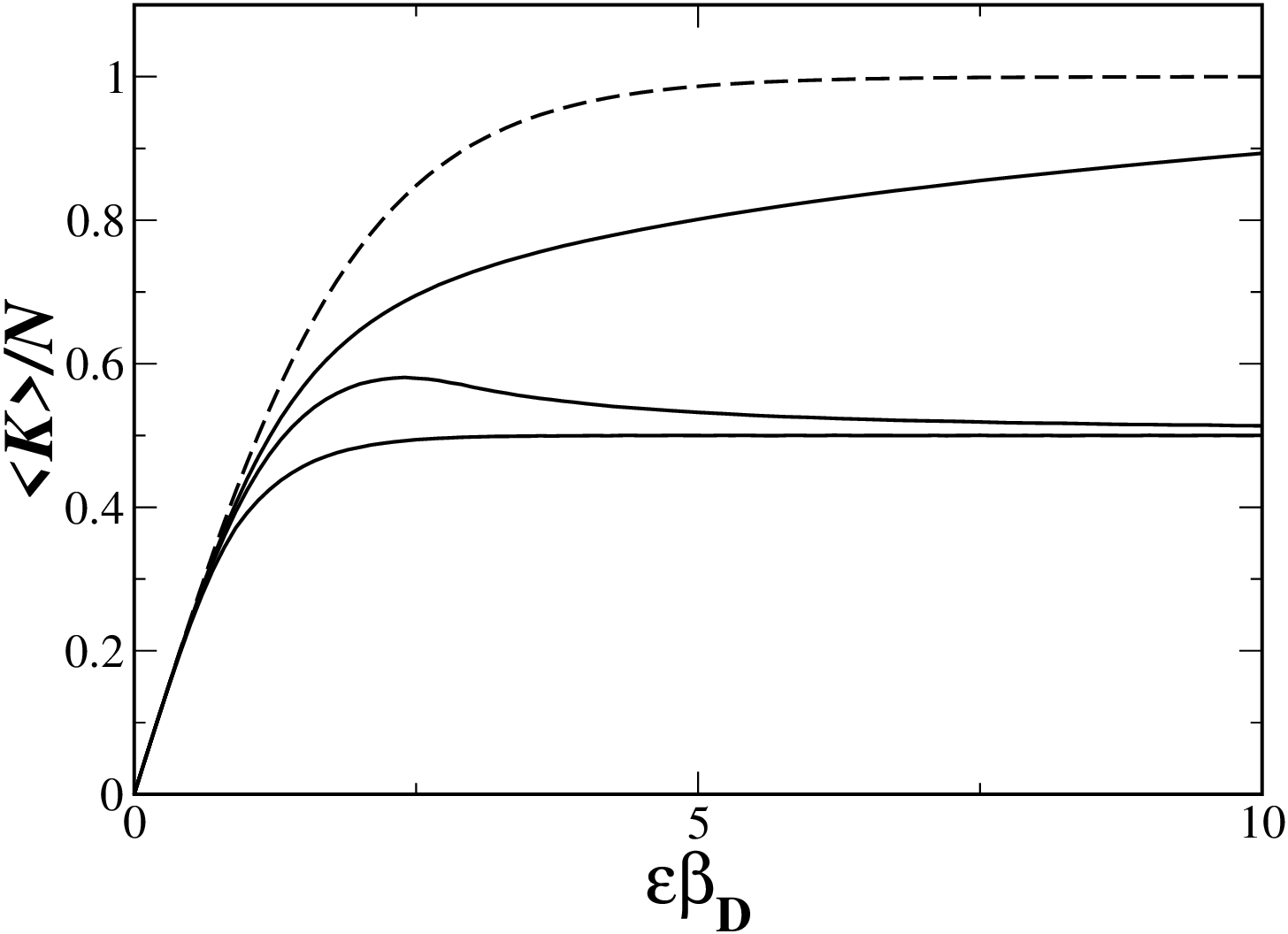}
\caption{\label{bild:anticoop_comp} Average complementarity of the probe ensemble for the
  anti-cooperative HP-model (\ref{eq:HP-coop}) with $J < 0$. The
  dashed line represents $J=0$, the solid curves from top to bottom
  correspond to the values -0.1, -0.2 and -0.5 of $J$ in units of
  $\varepsilon$.}
\end{center}
\end{figure}

Before analysing the recognition ability for $J\neq 0$ consider
briefly the influence of an anti-cooperative interaction with $J < 0$
in Hamiltonian (\ref{eq:HP-coop}) on the average complementarity
$\left < K\right>$.  From the discussion of the effect of the
cooperative term within the design step one may expect that
anti-cooperative interactions should decrease $\left < K \right>$. For
a probe molecule with a high complementarity to the target molecule
all $S_i$ tend to be in state $+1$ to ensure good contacts and thus a
large energy decrease due to the hydrophobic interaction.  However,
the anti-cooperative term then leads to an energy increase so that the
two contributions to the Hamiltonian (\ref{eq:HP-coop}) compete with
each other. Two different regimes can now be expected. Large values of
the parameter $\beta_{\mb{\tiny D}}$ favour structures $\theta$ that
are highly complementary to the target $\sigma^{(0)}$. For $0 > J >
-\varepsilon/8$ the hydrophobic interaction term is dominant leading
to a majority of good contacts $S_i = +1$ and thus $\left< K\right>$
is expected to become $N$ for increasing $\beta_{\mb{\tiny D}}$.
However, if $J < -\varepsilon/8$ the second anti-cooperative term
dominates leading to an alternating structure of good and bad contacts
where the $S_i$ of two neighbouring positions have different signs.
Note that in such a situation the direct hydrophobic-polar interaction
contributes a maximum favourable energy $-\varepsilon/2$ per site
whereas the cooperative term gives the maximum contribution $4J$ per
site giving the cross-over value $J=-\varepsilon/8$ for the considered
square geometry . For one half of the residues one therefore has
preferably good contacts so that the residue on the probe molecule is
of the same type as the one on the target molecule on average. For the
other half of positions, however, one has bad contacts. For those
positions the hydrophobic interaction term in (\ref{eq:HP-coop}) does
not contribute and the probabilities of the residue on $\theta$ to be
hydrophobic or polar at such positions are equal. For $J<
-\varepsilon/8$ one thus expects that $\left<K \right>$ tends to $N/2$
for increasing $\beta_{\mb{\tiny D}}$. These expectations are indeed
confirmed by numerical investigations as shown in figure
\ref{bild:anticoop_comp}.

\begin{figure}[h!]
\begin{center}
\includegraphics[scale=0.285,angle=0]{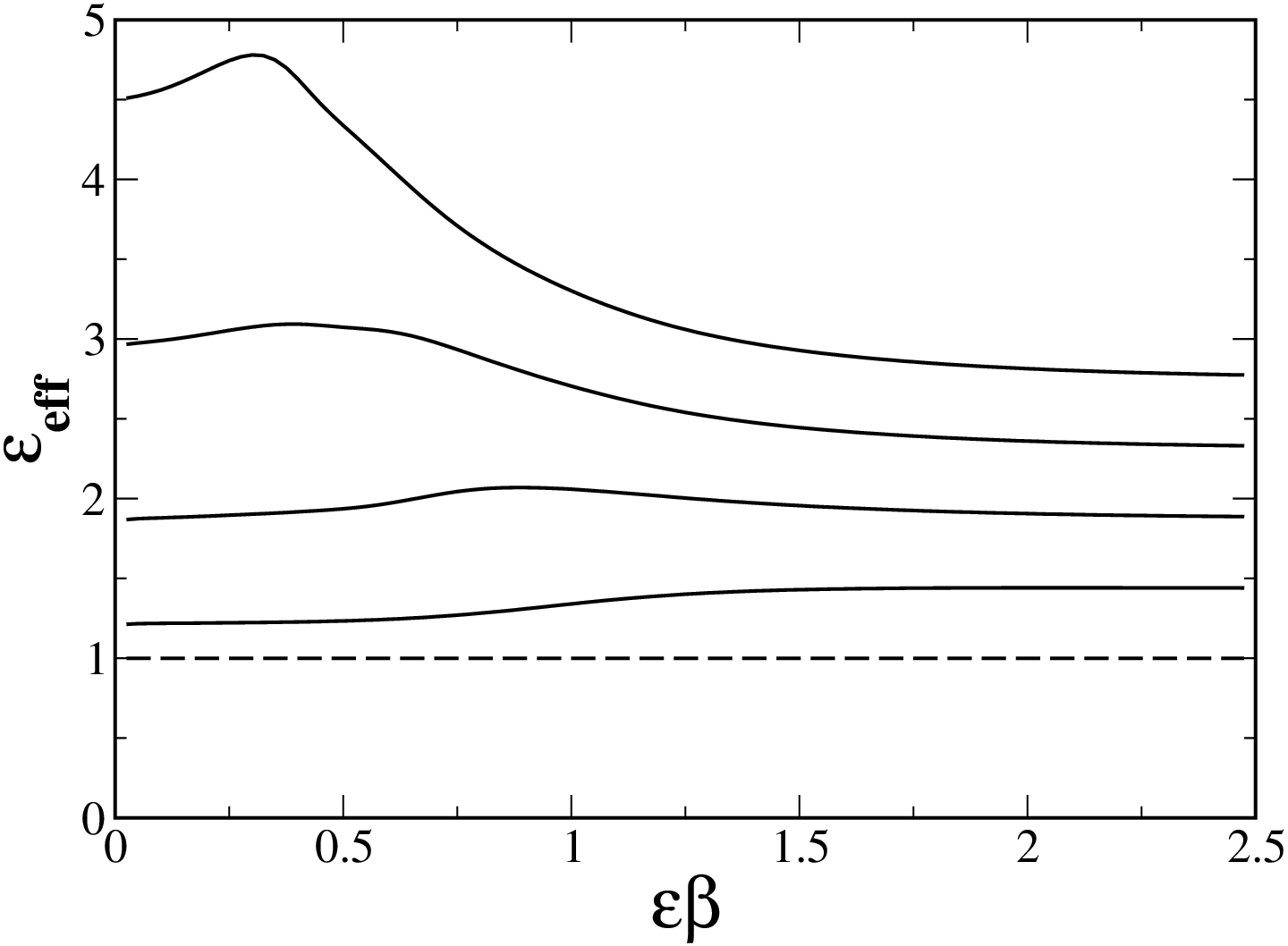}
\caption{\label{bild:espeff_comp} Effective interaction constant
  $\varepsilon_{\tiny\mb{eff}}$ as defined in the main text as a
  function of the temperature for the model (\ref{eq:HP-coop}). The
  solid curves correspond to values 0.25, 0.5, 0.75 and 1.0 of J (in
  units of $\varepsilon$) from the bottom up. }
\end{center}
\end{figure}

In the general discussion of the extended model (\ref{eq:HP-coop}) it
has been argued that the cooperative term will increase the effective
contribution of a residue-residue contact at the interface between the
two biomolecules. To get an impression of this increase one can define
an effective residue-residue interaction constant $\varepsilon_{\tiny
  \mb{eff}}(\beta_{\tiny\mb{D}},J) := \left<
  \mathcal{H}(J)\right>/\left< \mathcal{H}_{\mb{\sts HP}} \right>$
by considering the average interaction energy of the probe ensemble
with the target molecule for different values of the cooperativity
$J$. Figure \ref{bild:espeff_comp} shows that this effective
interaction constant is indeed increased by the cooperative term in
the Hamiltonian (\ref{eq:HP-coop}).

\paragraph{Recognition ability.}

\begin{figure}[h!]
\begin{center}
\includegraphics[scale=0.285,angle=0]{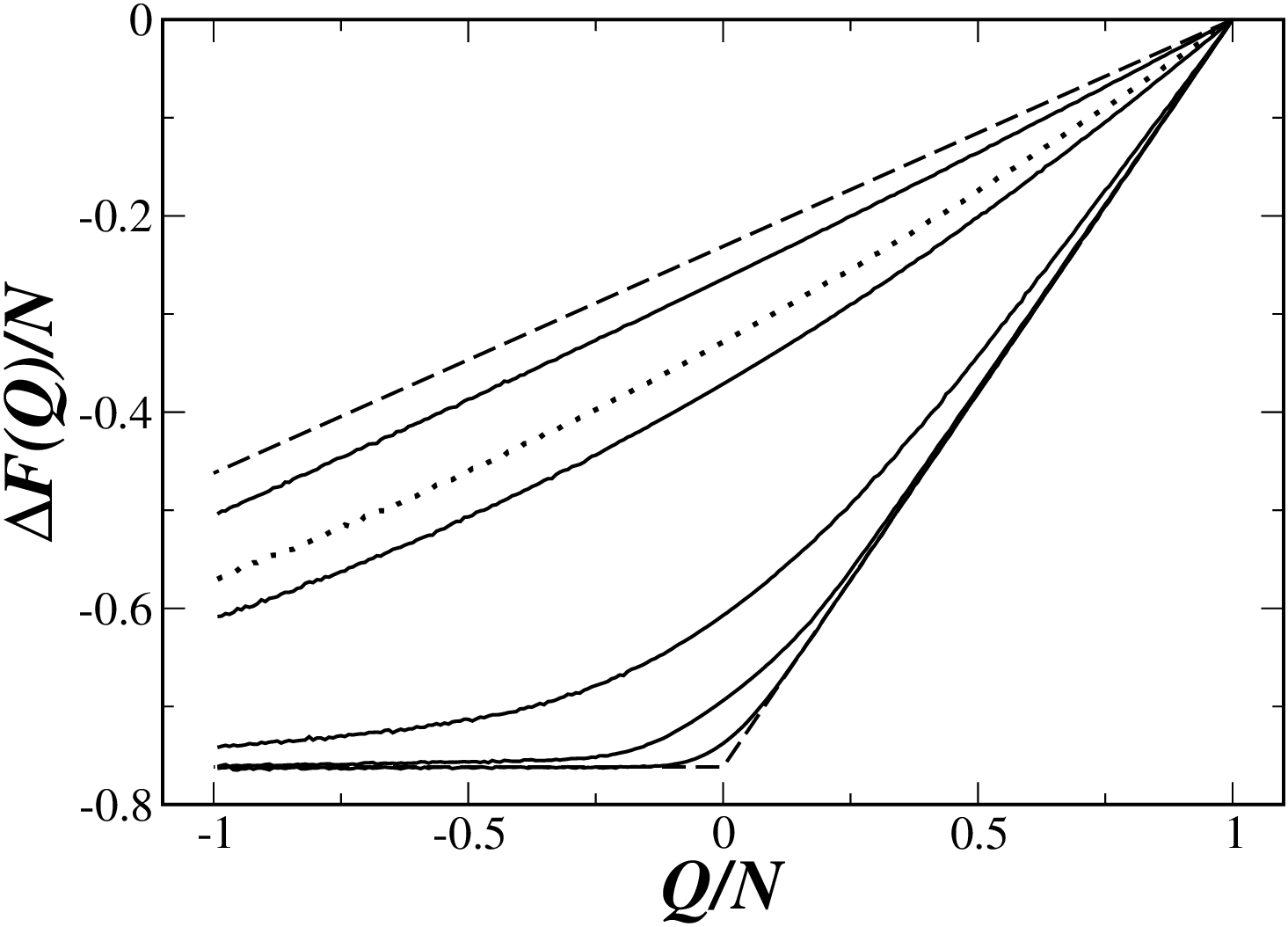}
\caption{\label{bild:coop_freiedifferenz} Free energy difference per
  site of the association of the probe ensemble with the two competing
  molecules as a function of their similarity for different
  cooperativities $J$ in (\ref{eq:HP-coop}). The upper dashed line
  corresponds to $J=0$. The lower dashed line describes the limiting
  case $J\to \infty$ in the limit of large $N$ (section
  \ref{kap:Jgross}). The solid curves from top to bottom correspond to
  the same values of $J$ as in figure \ref{bild:coop_comp}. The dotted
  curve shows the result for a system with additional next nearest
  neighbour cooperativity with $J_{\mb{\tiny nnn}}/\varepsilon =
  J_{\mb{\tiny nn}}/\varepsilon = 0.1$. The parameters
  $\beta_{\mb{\tiny D}}$ and $\beta$ are both 1.0. }
\end{center}
\end{figure}

The knowledge of the density of states allows the calculation of the
 recognition ability quantified by the free energy difference 
\begin{equation}
\label{eq:deltafvonQ}
\Delta F(Q) = \frac{\left< \delta_{Q,\sum_i\sigma_i^{(0)}\sigma_i^{(1)}} \Delta F(\sigma^{(0)},\sigma^{(1)})\right>_{\sigma^{(0)},\sigma^{(1)}}}{\left< \delta_{Q,\sum_i\sigma_i^{(0)}\sigma_i^{(1)}} \right>_{\sigma^{(0)},\sigma^{(1)}}}
\end{equation}
for the association of probe molecules with the two structures
$\sigma^{(0)}$ and $\sigma^{(1)}$. The results are shown in figure
\ref{bild:coop_freiedifferenz} for different values of the $J$.  For
comparison the free energy difference for the system with additional
next nearest neighbour cooperativity is shown as well.  An increase in
$J$ increases the free energy difference and therefore the recognition
specificity of the probe molecules. For a value of $J$ of the order
$\varepsilon$ the maximum effect of cooperativity has already been
reached for the considered temperature values $\beta_{\mb{\tiny D}} =
\beta = 1.0$. Thus, the expected increase of the recognition ability
by cooperativity for constants $J \simeq \varepsilon$ is indeed
confirmed by the numerical results.

\begin{figure}[h!]
\begin{center}
\includegraphics[scale=0.285,angle=0]{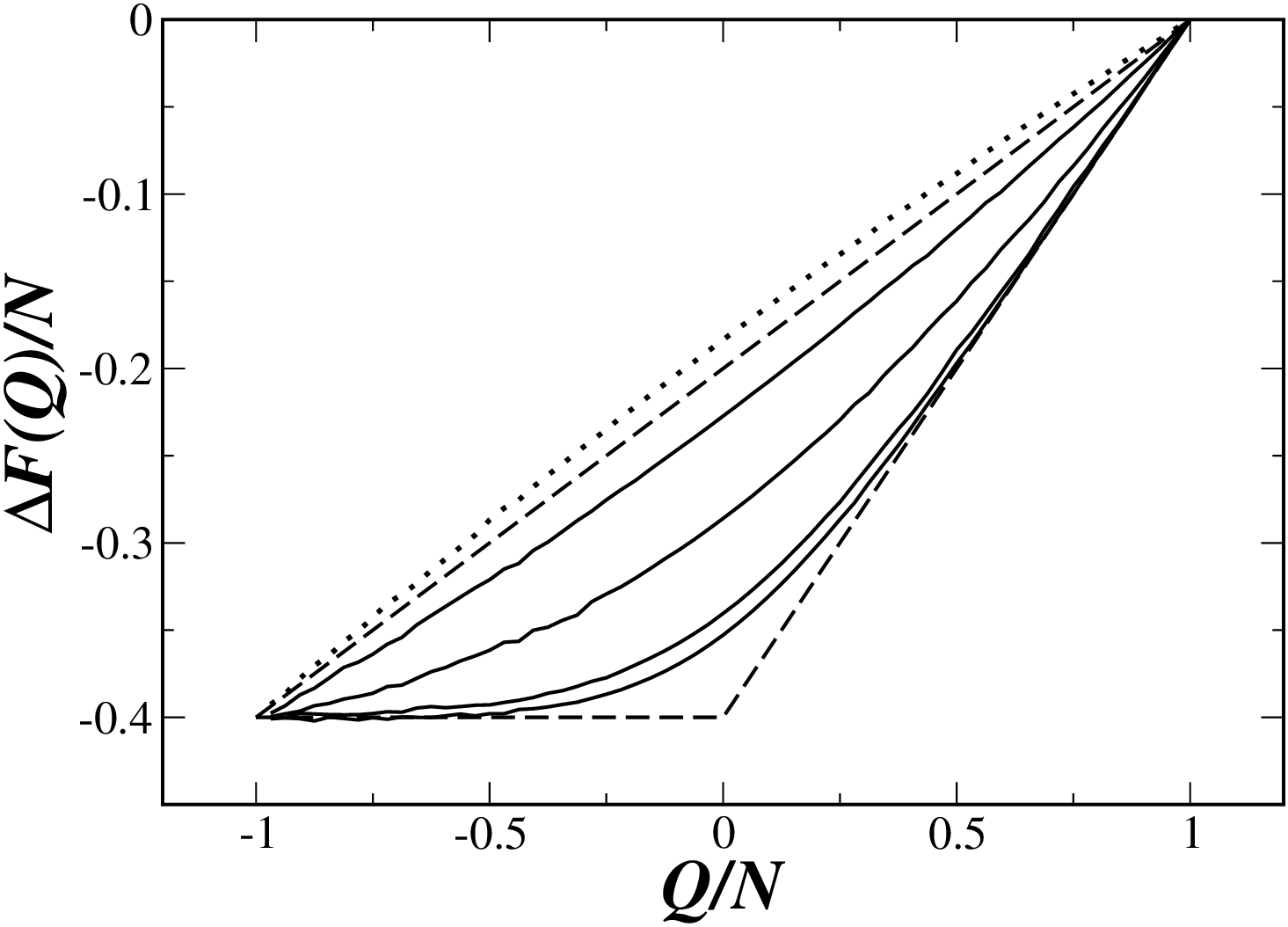}
\caption{\label{bild:coop_freiedifferenz_kompfest} Free energy
  difference as a function of the similarity for different
  cooperativities $J$ (with $\beta = 1.0$) where the probe ensemble
  has been designed to have a fixed $\left <K\right>/N = 0.4$. The
  upper dashed lines corresponds to $J = 0$, the lower one describes
  the limiting case $J \to \infty$ (and large $N$). The values of
  $J/\varepsilon$ in (\ref{eq:HP-coop}) are 0.25, 0.5, 0.75 and 1.0
  for the solid curves from top to bottom. For the dotted line
  $J/\varepsilon = -1/2$. }
\end{center}
\end{figure}

To study the influence of different cooperativities on the recognition
ability in a more direct way the following approach can be adopted.
The cooperativity already influences the design step and optimises the
probe ensemble with respect to the original target structure as can be
seen by the dependence of $\left < K\right >$ on the $J$. This better
optimisation influences the testing step as well. In order to
investigate the pure influence of the cooperative interaction on the
recognition ability more closely one can use probe ensembles where the
average complementarity is fixed to some values $K_0$ for different
$J$. This can be done by carrying out the design of the probe
molecules at different design temperatures such that $\left< K\right >
(\beta_{\mb{\tiny D}}, J) = K_0$. The probability distributions
obtained when this additional constraint is applied are then used to
calculated the difference of the free energy of association of the
probe molecules with both the target and the rival molecule. The
results are shown in figure \ref{bild:coop_freiedifferenz_kompfest}
for recognition sites with $N=64$ residues. Again it can be seen that
an increase in the cooperativity $J$ increases the free energy
difference for a fixed similarity $Q/N$ between the target and the
rival biomolecule.  The dashed lines in figure
\ref{bild:coop_freiedifferenz_kompfest} represent the free energy
difference for $J=0$ and for the asymptotic regime $J \to \infty$ with
$N\gg 1$. For large $Q/N$ and large $J$ the free energy difference is
already well represented by the asymptotic result. For a cooperativity
$J\simeq \varepsilon$ the maximum effect is thus already achieved.

For the minimum similarity parameter $Q = -N$ the free energy
difference at fixed $K_0$ is independent of the cooperativity $J$
(compare figure \ref{bild:coop_freiedifferenz_kompfest}).  To see this
consider the fixed structure $\sigma^{(0)}$ of the recognition site of
the target.  As the similarity parameter $Q$ is minimum, the competing
molecule has the structure $\sigma^{(1)} = -\sigma^{(0)}$ at its
recognition site.  The free energy difference of association is then
given by $\Delta F(-N) = -\frac{1}{\beta}\sum_\theta
P_J(\theta|\sigma^{(0)})(\ln Z(\theta|\sigma^{(0)})- \ln
Z(\theta|-\sigma^{(0)})) $. The partition function
$Z(\theta|\sigma^{(1)}) = Z(\theta|-\sigma^{(0)})$ related to the
rival structure explicitly reads
\begin{eqnarray}
  Z(\theta| - \sigma^{(0)})  =
  \mb{e}^{-\frac{\beta\varepsilon}{2}\sum_i\sigma_i^{(0)}
    \theta_i}
\\ \qquad\quad \times 
\sum_S \mb{e}^{-\beta\varepsilon\sum_{i} \frac{S_i}{2} \sigma_i^{(0)} \theta_i +\beta J
\sum_{\left<ij\right>} S_i S_j  }\\ \qquad =
 \mb{e}^{-{\beta\varepsilon}\sum_i\sigma_i^{(0)}
  \theta_i}Z(\theta|\sigma^{(0)}).
\end{eqnarray}  
where a transformation $S_i \to -\tilde{S}_i$ has been used for the
last equality. Note that the phase space for $\tilde{S}$ is the same
as for the variable $S$. Thus the free energy difference at $Q = -N$
is generally given by
\begin{equation}
  \Delta F(Q = -N) = -\varepsilon \sum_K P_{J}(K;\beta_{\mb{\tiny D}}) K =-\varepsilon
  \left<K\right>(J).
\end{equation}
As the average $\left<K\right>$ is fixed to the value $K_0$ for
different $J$ the free energy difference is the same for all $J$.

\begin{figure}
\begin{center}
\includegraphics[scale=0.285,angle=0]{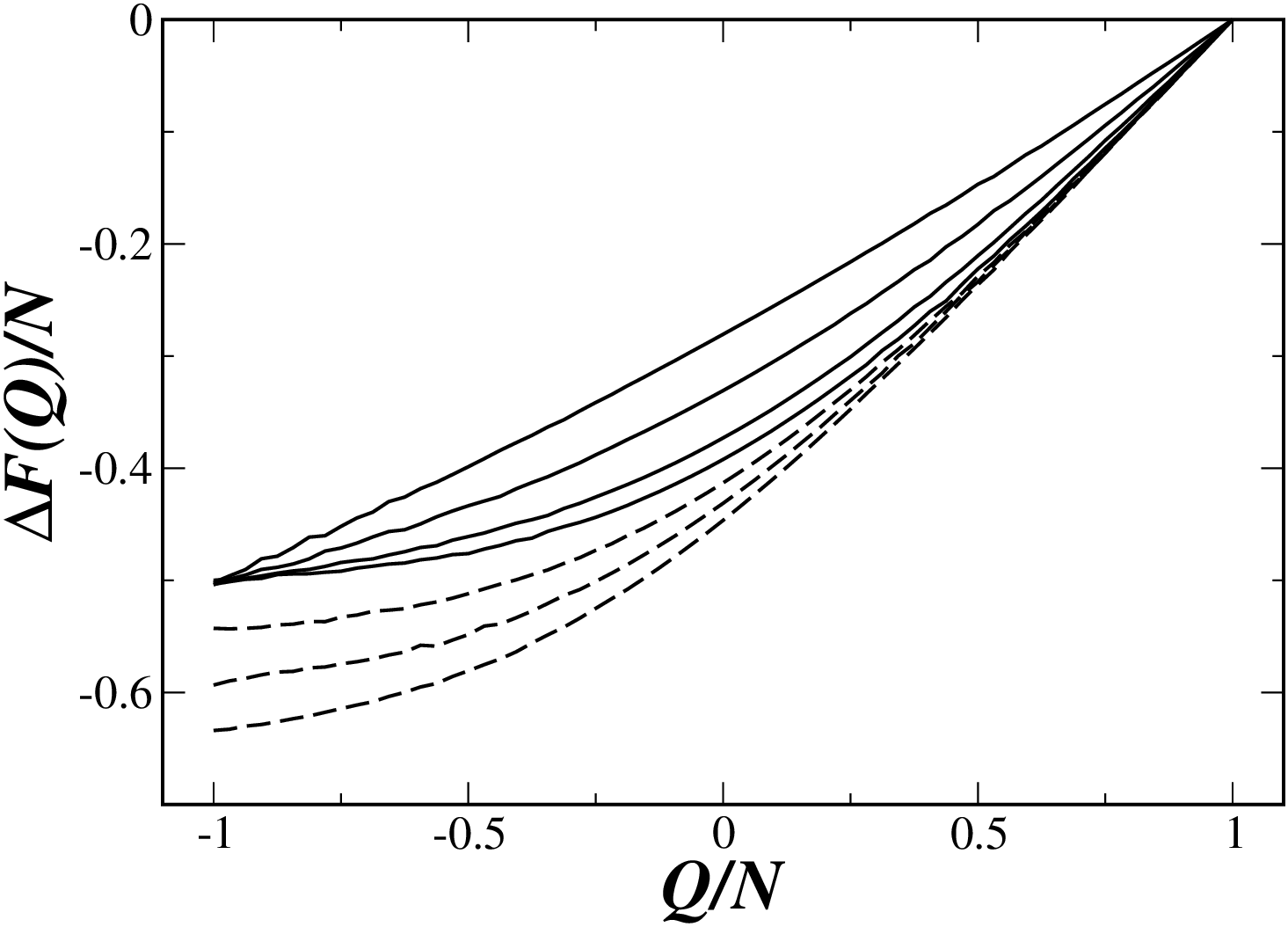}
\caption{\label{bild:coop_freiedifferenz_temp} Free energy difference as a function of the similarity for fixed 
  $J/\varepsilon = 1/2$ in (\ref{eq:HP-coop}) and different testing
  temperatures $\beta$. The probe ensemble has been designed to have
  $\left<K\right> /N = 0.5$. The solid curves correspond to the $\beta
  = 0.5, 0.75, 1.0$ and $1.25$ from top to bottom. For the dashed
  curves $\beta = 1.25$ and an additional field $\gamma$ has been
  applied, namely from top to bottom $\gamma/\varepsilon = 0.05, 0.1$
  and $0.15$.  }
\end{center}
\end{figure}

For the HP-model with pure hydrophobic interactions the free energy
difference is independent of the conditions under which the
recognition ability is tested. It is only determined by the design
conditions (compare relation (\ref{eq:deltafreinhp})). For the
extended HP-model (\ref{eq:HP-coop}) with cooperative
interactions this is no longer the case. Apart from the design
conditions, encoded in the Lagrange parameter $\beta_{\mb{\tiny D}}$,
the free energy difference depends on the $\beta$ which
specifies the conditions for the testing step. In figure
\ref{bild:coop_freiedifferenz_temp} the free energy difference is
shown for different values of $\beta$. The cooperativity
constant is fixed to be $J/\varepsilon = 1/2$, the design temperature
$\beta_{\mb{\tiny D}}$ is chosen to have $\left< K\right> = N/2$. For
increasing parameters $\beta$ the absolute value of the free energy
difference is increased.
For the minimum similarity $Q = -N$ the free energy difference becomes
independent of $\beta$. Its value at the minimum similarity is only
determined by the design conditions and is given by $\Delta F (Q = -N)
= -\varepsilon \left<K\right>$ as has been shown above.

The independence of $\Delta F (Q = -N)$ of the testing temperature
$\beta$ is a result of the symmetry of the underlying model
(\ref{eq:HP-coop}). This symmetry is broken by introducing a field
like term $-\sum_i\gamma S_i$ to the energy. It is expected that there
is some bias towards good or bad contacts leading to a such an
additional field with $\gamma \neq 0$.  For positive fields $\gamma$
the recognition ability is again expected to be increased with respect
to the situation where $\gamma$ vanishes. This is shown by the dotted
lines in figure \ref{bild:coop_freiedifferenz_temp}.

\subsection{Mean field theory for arbitrary cooperativity}
\label{kap:meanfieldJ}

After having analysed the influence of the cooperative terms in the
previous paragraphs by means of an asymptotic analysis and Monte Carlo
simulations we briefly sketch how a mean-field treatment can be
carried out \cite{Behringer_zukunft}. The discussion will be
restricted to the determination of the averaged complementarity. As
already mentioned the variable $\sigma_i\theta_i$ acts as a random
field in (\ref{eq:HP-coop}) and therefore techniques from the theory
of disordered systems have to be applied in the mean field treatment
(see, for example, \cite{Dotsenko_2001,Schneider_1977}). Thus the
auxiliary variable $k_i = \theta_i\sigma^{(0)}_i$ which has been
introduced in \ref{kap:Jgross} and specifies a complementarity
configuration $k = (k_1, \ldots,k_N)$ is used in the following.  The
mean-field approach consists of two steps, namely an equivalent
neighbour approximation of the cooperative interaction term and an
asymptotic evaluation of the partition sum for large $N$. The
equivalent neighbour approximation of the Hamiltonian
(\ref{eq:HP-coop}) reads
\begin{equation}
  \label{eq:en-approx}
\mathcal{H}_{\mb{\sts EN}} = -\frac{J}{2N}\left(\sum_i S_i\right)^2 -
{\varepsilon}\sum_i \frac{1 + S_i}{2} k_i.
\end{equation}  
We aim at a calculation of the averaged complementarity $K = \left
  <\sum_i k_i \right>$ containing a thermal average with respect to the
interaction variable $S$ and an average over the possible
complementarity configurations $k$ of the probe molecules with respect
to the target. The thermal averaged leads to the distribution
$P(\theta|\sigma^{(0)})$ of probe molecules and thus to a distribution
$P(k)$ of the complementarity configuration itself. Consider first the
thermal average with respect to $S$.  The variable $x:= \sum_i S_i$
appears quadratically in (\ref{eq:en-approx}). By introducing an
additional auxiliary variable $y$ it can be linearised in the argument
of the Boltzmann factor in the partition sum $Z(k) = \sum_S
\exp(-\beta \mathcal{H}_{\mb{\sts EN}})$ with the help of the identity
\begin{equation}
  \exp\left(\frac{a}{2N} x^2\right) = \int\limits_{-\infty}^{+\infty} \mb{d} y \sqrt{\frac{Na}{2 \pi}}
  \exp\left(-\frac{Na}{2}y^2 + axy\right),
\end{equation}
often called Hubbard-Stratonovich transformation in the literature.
Note that the distribution function $P(k)$ of the complementarity
configuration is determined by $Z(k)$ up to the normalisation. The
summation over $S$ can then be carried out leading to
\begin{equation}
\label{eq:Z-integral}
  Z(k) \sim \exp\left(\frac{\beta\varepsilon}{2}\sum_i k_i\right) \int\limits_{-\infty}^{+\infty} \mb{d} y
  \exp\left( \mathcal{A}(y,k) \right)
\end{equation}  
with
\begin{equation}
\mathcal{A}(y,k) = - \frac{\beta JN}{2}y^2 + \sum_i \ln \cosh\left( \beta J y
  +\frac{\beta\varepsilon}{2} k_i\right).
\end{equation}
In the asymptotic limit of large $N$ the integration over the
auxiliary field $y$ in (\ref{eq:Z-integral}) can be carried out using
the Laplace method (e.\,g. \cite{Jaenich2001,Murry1984}). This gives
\begin{equation}
  Z(k) \sim \exp\left( \frac{\beta\varepsilon}{2}\sum_i k_i + \mathcal{A}(y_0,k)\right)
\end{equation}  
aside from irrelevant factors.  The mean field $y_0$ is determined by
the saddle point equation
\begin{equation}
  y_0 = \frac{1}{N} \sum_i \tanh\left( \beta J y_0 +\frac{\beta\varepsilon}{2}k_i\right).
\end{equation}
Note that the mean field explicitly depends on the local
complementarity configuration $k$. These two equations can be used to
carry out the configurational average over all $k$ to obtain the
averaged complementarity $\left<K \right>$ by noting that a particular
configuration $k$ contains $K^{(+)}$ sites with $k_i = +1$ and
$K^{(-)}$ ones with $k_i = -1$. The partition function $Z$ (and thus
the distribution function $P$) as well as the mean field $y_0$ are
therefore only functions of $K^{(\pm)}$. The mean field $y_0(K^{(+)},
K^{(-)})$, for example, is then given by
\begin{equation}
  y_0
 = \frac{K^{(+)}}{N} \tanh\left( \beta J y_0
 +\frac{\beta\varepsilon}{2}\right) +  \frac{K^{(-)}}{N} \tanh\left(
 \beta J y_0 -\frac{\beta\varepsilon}{2}\right).
\end{equation}
The average over $k$ can thus be converted to an average over
$(K^{(+)},K^{(-)})$ so that the complementarity $\left<K \right> =
\left< K^{(+)} - K^{(-)}\right>$ can by worked out using a computer
algebra program. The results are shown in the inset of figure
\ref{bild:coop_comp} together with the Monte Carlo findings discussed
in the previous paragraph. The mean field curves behave qualitatively
similar as the Monte Carlo curves. Using a similar decomposition of
the similarity configuration $q_i = \sigma^{(0)}_i\sigma^{(1)}_i$
between the target and the rival structure into positive contributions
$Q^{(+)}$ and negative ones $Q^{(-)}$ one can calculate the averaged
free energy difference $\Delta F(Q)$ (compare relation
(\ref{eq:deltafvonQ})). The resulting curves show again the same
qualitative behaviour as the results from the Monte Carlo simulations.

\section{Cooperativity coupling to residue structure}
\label{kap:alter_coop}

The importance of cooperativity in biological situations was
emphasised at the beginning of section \ref{kap:coop}.  In Hamiltonian
(\ref{eq:HP-coop}) an additional cooperative term has been introduced
which, however, does not couple to the residue distributions on the
recognition sites of the two molecules in contact. In general the
additional cooperative interaction terms might also couple to the
structures $\sigma$ and $\theta$ of the target and probe molecule,
respectively. One possible coupling is given by the Hamiltonian
\begin{equation}
\label{eq:HP-alt-coop}
\mathcal{H}(\sigma, \theta;S) = - \sum_{i=1}^{N}\left(\varepsilon \frac{1+S_i}{2}  + J
\sum_{i_\delta} S_i S_{i_\delta}\right)\sigma_i \theta_i.
\end{equation}
The sum in the second term extends over the neighbouring positions
$i_\delta$ of the position $i$ on the interface. Again the cooperative
term will lead to an additional energy contribution depending on how
the side chains are rearranged in the interface. In case of a
favourable direct energy contribution from the hydrophobic interaction
at site $i$ described by the product $\sigma_i\theta_i$ the
cooperative term rewards good contacts like in the Hamiltonian
(\ref{eq:HP-coop}). However, in (\ref{eq:HP-coop}) two neighbouring
bad contacts due to an unfavourable hydrophobic-polar interaction are
also attributed a favourable cooperative contribution. This is no
longer the case in Hamiltonian (\ref{eq:HP-alt-coop}) as the sign of
the cooperative energy contribution now depends on the sign of the
hydrophobic interaction energy at position $i$ on the interface. It is
thus expected that the cooperative terms in (\ref{eq:HP-alt-coop})
lead to a more favourable cooperative contribution than those in
Hamiltonian (\ref{eq:HP-coop}). The cooperative terms in the
Hamiltonians (\ref{eq:HP-alt-coop}) and (\ref{eq:HP-alt-coop}) are
only two possible ways to take into account cooperativity, corresponding in our
modelling to mutual interaction of neighbouring variables $S_i$, other extensions are possible as well.

As already remarked the variable $\sigma_i\theta_i$ in
(\ref{eq:HP-coop}) is basically a random field the distribution of
which is determined by the design step. The model (\ref{eq:HP-coop})
is thus a random field Ising model where the random field
$\sigma_i\theta_i$ is asymmetrically distributed. In Hamiltonian
(\ref{eq:HP-alt-coop}) this random variable now also couples to the
exchange constant $J$ of the interactions between neighbouring
variables $S_i$ and thus the exchange constant also becomes a random
variable. The model (\ref{eq:HP-alt-coop}) is thus an
Edward-Anderson-like model in a random field with an asymmetrically
distributed exchange constant $J\sigma_i\theta_i$.

\begin{figure}[h!]
\begin{center}
\includegraphics[scale=0.285,angle=0]{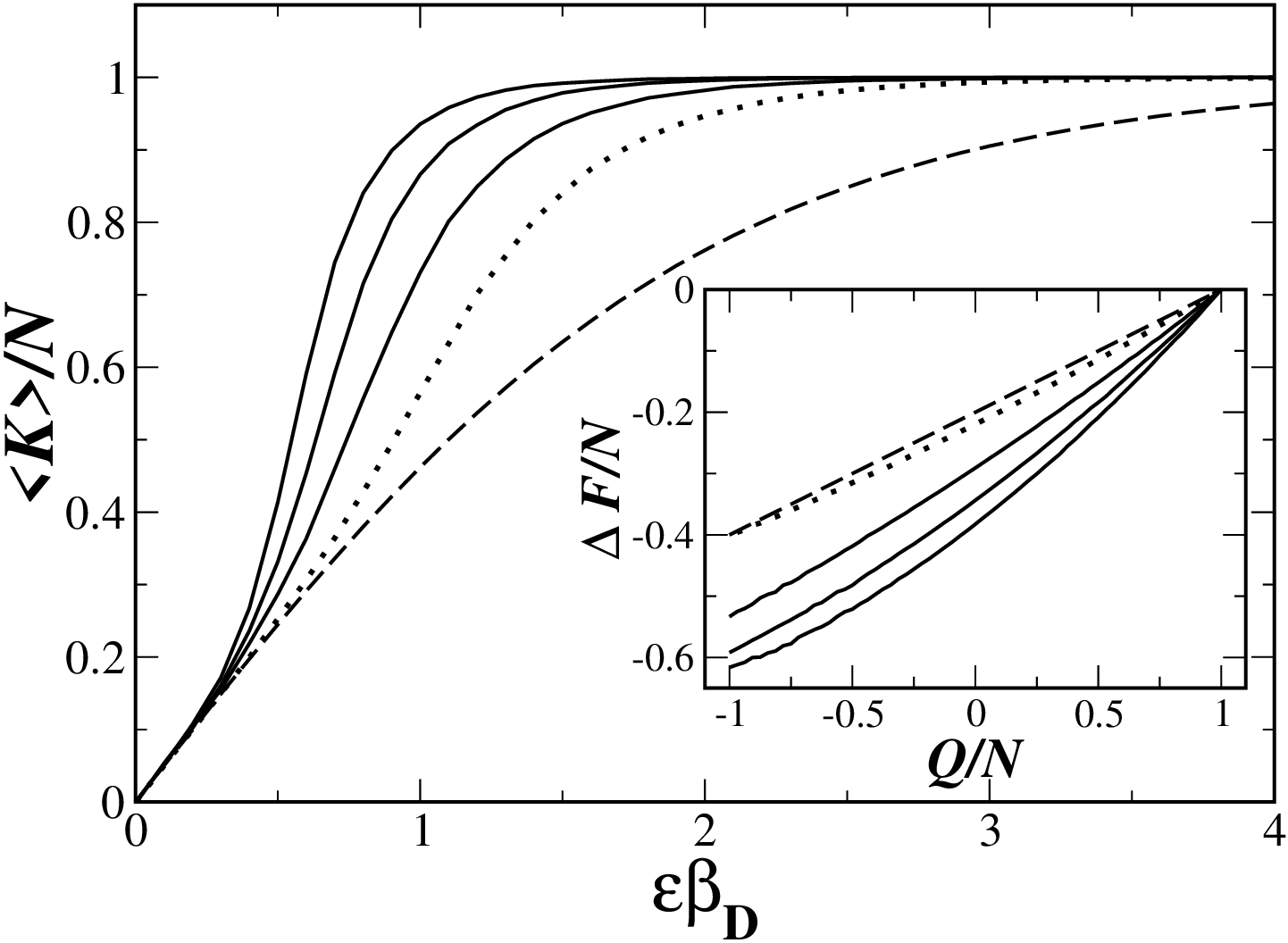}
\caption{\label{bild:coop_alternativ} 
  Averaged complementarity of the probe ensemble designed according to
  the model (\ref{eq:HP-alt-coop}). The dashed curve represents $J =
  0$, for the solid curves $J/\varepsilon = 0.2, 0.3$ and $0.4$ from
  bottom up. For comparison, the dotted line depicts the corresponding
  curve for the model (\ref{eq:HP-coop}) with $J/\varepsilon = 0.2$.
  The inset shows free energy differences as a function of the
  similarity $Q$ for the same set of parameters, where
  $\beta_{\mb{\tiny D}}$ is chosen to have $\left<K\right>/N = 0.4$
  (with $\beta = 1.0$ for each curve).  }
\end{center}
\end{figure}

The two stage approach to obtain the recognition ability can now be
carried out numerically for the model (\ref{eq:HP-alt-coop}) by
calculating again density of states $\Omega_J(K;E)$ by a Monte Carlo
simulation.  The results for the averaged complementarity of the probe
molecules and the free energy difference are depicted in figure
\ref{bild:coop_alternativ}. One observes a similar qualitative
behaviour as the corresponding curves for the model
(\ref{eq:HP-coop}). Again, it is found that an increase of the
parameter $J$ increases the quality of the design step in the sense
that the probe molecules are better optimised with respect to the
target biomolecule indicated by an increase of $\left <K\right>$ for
higher values of $J$.  Similarly, the recognition ability measured by
the free energy difference $\Delta F = F^{(0)} - F^{(1)}$ for a given
similarity $Q$ between the target and the rival grows for increasing
$J$.

\section{Conclusions and outlook}
\label{kap:conclusion}

We have presented coarse-grained models to study properties of
molecular recognition processes between rigid biomolecules. The
development of the models has been motivated by
experimental investigations on the biochemical
structure of the interface of protein complexes. A two-stage
approach containing a design of probe molecules and a testing of
their recognition ability has been adopted. This approach has been
used to investigate the role of cooperativity in molecular
recognition. The coarse-grained models capture the effects of
cooperativity on a residue specific level. The necessity of such an
approach has been pointed out in the literature \cite{Cera_1998}. We
have 
shown numerical results and compared them to analytic results obtained
in the asymptotic limit where cooperative interactions dominate over
direct hydrophobic interactions between the residues and in the
mean-field formulation of the models. It turned out that a small
contribution due to cooperativity can already substantially influence
the recognition ability, corroborating the suggestion
that cooperativity has a considerable effect on the
recognition specificity. Two possibilities to include
cooperative interactions have been explicitly analysed leading to
similar qualitative results. We note in passing that the proposed coarse-grained
model can reproduce qualitatively the experimental observation that in
antigen-antibody complexes, which require a relatively high binding
flexibility, a small number of strong non-covalent bonds across the
interface seems to be favoured compared to a situation with many but
rather weak bonds. The details are published elsewhere
\cite{Behringer_etal_2006}.

The proposed approach to study molecular recognition with
coarse-grained lattice models can be
extended in various ways. Apart from working with refined models, which
capture more details of the actual physical interactions across the
interface of the two biomolecules, the design step can be modified to
mimic natural evolution in a more realistic
manner. The presented analysis considered on the level of the target
and the rival molecule is basically a single molecule approach,
although the molecules are described in a very coarse-grained way.  The
influence of the heterogeneity of the mixture of target and rival
molecules encountered in real physiological situations as found in a
cell, for example, can be incorporated in our analysis. To this end
ensembles of targets and rivals differing in certain properties as for
example correlations and length scales have to be considered. A recent
study indeed indicates that the local small-scale structure related to
the distribution of the hydrophobicity on the recognition site of the
biomolecules seems to play a crucial role in molecular
recognition \cite{Bogner_2004}.

\acknowledgments

We thank the Deutsche Forschungsgemeinschaft (SFB 613) for financial support.



\begin{thebibliography}{66}
\bibitem{Behringer_zukunft} We are currently investigating further
  physical aspects of the introduced coarse-grained models for
  molecular recognition with mean-field methods. More details of the
  mean-field calculations will be published in a corresponding
  article.
\end{thebibliography}
\end{document}